\documentclass[11pt]{article}

\usepackage[T1]{fontenc}
\usepackage{lmodern}
\usepackage[margin=1in]{geometry}
\usepackage{amsmath,amssymb}
\usepackage{graphicx}
\usepackage{microtype}
\usepackage{xcolor}
\usepackage[round]{natbib}
\usepackage[colorlinks,allcolors=blue!50!black]{hyperref}

\graphicspath{{figs/}}

\newcommand{\prefcent}{\texttt{prefcent}}

\title{Preferential centrality with a density penalty:\\
       formulation and an open-source implementation}
\author{Alexander Hellervik\thanks{Chalmers University of Technology.
        \texttt{alexander.hellervik@chalmers.se}}}
\date{5 October 2026}

\begin{document}
\maketitle

\begin{abstract}
Preferential centrality \citep{hellervik2019} models spatial activity through
feedback between activity and attraction. We extend it with a density penalty. The
local response depends on activity per unit of capacity and scales with capacity. A
correction that is uniform per unit of capacity preserves total activity, and for a
chosen response function and penalty strength, conservation determines this
correction. We provide \prefcent, an open-source reference implementation with
convergence and feasibility checks and records of each run's inputs and settings.
It reproduces the 200-iteration outputs of a companion Porto Alegre study across
all $65{,}357$ network segments, with median relative differences of order
$10^{-14}$. In a strong-feedback example, the penalty reduces concentration and
reaches the stopping tolerance in fewer iterations. Other synthetic examples show
how penalty strength changes the relative prominence of centers and how their
number varies with transport costs. A further example reaches different numerical
equilibria from different initial states on the same network and capacity
landscape. In multi-start sweeps on three synthetic two-dimensional landscapes,
distinct converged states occur over a range of parameter values: at 172 of 420
points of a grid in feedback strength and transport cost, including 75 where the
states differ in their number of centers. The note provides a model specification and a reproducible
computational basis for further theoretical and empirical work.
\end{abstract}

\section{Introduction}
\label{sec:intro}

Preferential centrality answers a simple question: given a transport network and a
map of where activity can locate, where does it locate? The model takes a network
with generalized travel costs and a static capacity landscape as input, and returns
an equilibrium distribution of activity as output. Its defining ingredient is that
attraction is partly endogenous: a location attracts interaction in proportion to
what the land offers plus what activity is already there. \citet{hellervik2019}
introduced preferential centrality as an extension of spatial-interaction models in
which the flows leaving each origin are fixed by its activity, the
production-constrained models of \citet{wilson1970} and \citet{harriswilson1978}.
\citet[Sections 1.2 and 8.4--8.5]{hellervik2021}
develops the broader conceptual framework linking activity, interaction and
transport networks.

In the strong-feedback example of Figure~\ref{fig:rescue}, the unpenalized model
concentrates activity in a narrow region and approaches that state slowly. We add
a density penalty to reduce this concentration. The penalty responds to local
activity relative to capacity without specifying the mechanism, which may be
congestion, rents, land-use regulation or capacity limits. Travel costs remain
fixed. \citet[Section 12.4]{hellervik2021} identified
congestion feedback between activity, network flow and travel times as a direction
for further development; the present extension uses a local density response
rather than an explicit model of those feedbacks.

Congestion and destination crowding have previously been incorporated into
spatial-interaction models, including the destination-choice game of
\citet{yanzhou2019}. Here the density response enters as an additive correction to
preferential-centrality inflow, rather than as a crowding cost in the
destination-choice utility. The correction subtracts from each zone's density cost its
capacity-weighted average, so the adjustments sum to zero; evolutionary game dynamics
use the same construction to keep population shares summing to one
\citep[eq.~3.24]{mertikopoulos2018}. The present
contribution is this density-penalized preferential-centrality specification and its
reference implementation.

Section~\ref{sec:model} specifies the density penalty and gives an equilibrium
interpretation. Section~\ref{sec:solver} describes the reference implementation and
reproduces the outputs of the Porto Alegre computation. Section~\ref{sec:illustrations}
presents synthetic examples, including multi-start sweeps across parameter grids, and
Section~\ref{sec:enables} discusses possible applications to settlement history.

\section{The model and the density penalty}
\label{sec:model}

\subsection{The base model}
\label{sec:base}

There are $N$ zones on a transport network with generalized costs $c_{ij}$ between
them. Each zone $j$ has a fixed capacity $R_j > 0$ (buildable area, segment length,
or similar) and a nonnegative activity mass $a_j$. Capacity sets the scale of
activity; it is not a hard upper bound on $a_j$. Symbols without a zone index
denote vectors over the zones, for example $a = (a_1, \dots, a_N)$ and $R$, and $K$
denotes the matrix of the $K_{ij}$ defined below. A map of the state, such as the
inflow $I(a)$ or the update $C(a)$ of Section~\ref{sec:penalty}, returns a vector, and
$C_j(a)$ is its component at zone $j$. Interaction follows a
production-constrained gravity model: each unit of activity at zone $i$ generates
interaction, distributed over destinations $j$ in proportion to attraction times a
distance-decay kernel,
\begin{equation}
  S_{ij} \;=\; a_i \, \frac{W_j\, K_{ij}}{\sum_k W_k\, K_{ik}},
  \qquad K_{ij} = m_{ij}\,(c_{ij} + d_0)^{-\beta},
  \label{eq:gravity}
\end{equation}
where the mask $m_{ij}$ is $0$ for $i = j$ and for pairs with no route between
them, and $1$ otherwise. All runs in this note exclude self-interaction; the
manifest records this setting. The parameter $d_0$ is a fixed access cost added to
every trip. The attraction is
\begin{equation}
  W_j \;=\; \gamma\, a_j + R_j ,
  \label{eq:attraction}
\end{equation}
where $\gamma \ge 0$ sets how much attraction is generated by activity itself.%
\footnote{\citet{hellervik2019} write the attraction as $a_j + \alpha R_j$; for
$\gamma > 0$ this is equation~(\ref{eq:attraction}) with $\gamma = 1/\alpha$, because
a common factor in $W$ cancels in equation~(\ref{eq:gravity}).}
Equilibrium closes the loop: activity at a zone equals the interaction it receives,
\begin{equation}
  a_j \;=\; \sum_i S_{ij},
  \label{eq:balance}
\end{equation}
with total activity fixed at $\sum_j R_j$. At $\gamma = 0$ the map is linear and
the equilibrium is a type of eigenvector centrality \citep{hellervik2019}; at $\gamma > 0$ activity and
attraction reinforce each other.

\subsection{The density penalty}
\label{sec:penalty}

Write $I_j(a) = \sum_i S_{ij}(a)$ for the inflow at the current state and
$\tilde a_j = a_j / R_j$ for density. Equation~(\ref{eq:balance}) is the unpenalized
balance equation $a = I(a)$. We modify it to $a = C(a)$ so that density pushes
back on concentration, imposing three structural requirements and then specifying a
redistribution convention.

\paragraph{R1 (local, per unit of capacity).} The local response depends on
density and scales with capacity. We write it as $\kappa\, R_j\, g(\tilde a_j)$, with a
strength $\kappa \ge 0$ and a response function $g$. A separate correction
preserves total activity.

\paragraph{R2 (subdivision consistency).} Splitting a zone into two co-located
halves leaves the aggregate dynamics unchanged. Each half has half the original
capacity and activity. The copies inherit the original kernel relationships and do
not interact with each other. This condition does not cover repartitioning that
changes locations or travel costs.

\paragraph{R3 (conservation).} Total activity is fixed at $\sum_j R_j$, so the
adjustments must sum to zero at every state, not only at equilibrium.

\paragraph{Redistribution convention.} R1--R3 allow several redistribution
rules.\footnote{For example, returning the removed penalty in proportion to
$R_j\,h(\tilde a_j)$ for a positive $h$ also conserves total activity and scales
correctly under co-located splitting; $h = 1$ is the convention used here, and
$h(\tilde a) = \tilde a$ would return it in proportion to activity. The convention is
therefore an assumption, not a consequence of R1--R3.} We choose a correction that is uniform per unit of capacity. Write
\begin{equation*}
  C_j(a) \;=\; I_j(a) \;-\; \kappa\, R_j\, g(\tilde a_j) \;+\; R_j\, b(a),
\end{equation*}
where $b(a)$ is the same for all active zones. Conservation gives
\begin{equation*}
  b(a) \;=\; \kappa\, \frac{\sum_k R_k\, g(\tilde a_k)}{\sum_k R_k} \;=\; \kappa\,\bar g_R .
\end{equation*}
Substitution gives
\begin{equation}
  C_j(a) \;=\; I_j(a) \;-\; \kappa\, R_j \bigl(g(\tilde a_j) - \bar g_R\bigr).
  \label{eq:penalty}
\end{equation}
For fixed $g$ and $\kappa$, the chosen redistribution convention determines this
correction. For $\kappa > 0$, the correction lowers a zone's update relative to
$I_j$ when $g(\tilde a_j) > \bar g_R$ and raises it when $g(\tilde a_j) < \bar g_R$; the
adjustments sum to zero. All sums run over the active zones defined in
Section~\ref{sec:failstop}. Subtracting an unweighted mean of $g$ instead would
violate R2: the unweighted mean counts zones, so subdividing one zone would change
the correction everywhere. Given the convention, the capacity weighting of the mean
is a consequence of R2 and R3.

\paragraph{Equilibrium interpretation.} The same expression has an equilibrium
interpretation. Call $(I_j - a_j)/R_j$ a zone's net surplus per unit of capacity,
and let density cost a zone $\kappa\, g(\tilde a_j)$ per unit of capacity. Suppose the
distribution settles where no zone differs in surplus net of density cost,
\begin{equation}
  \frac{I_j - a_j}{R_j} - \kappa\, g(\tilde a_j) \;=\; c,
  \qquad \text{the same $c$ for all active zones.}
  \label{eq:equalization}
\end{equation}
Multiplying by $R_j$ and summing, conservation ($\sum_j a_j = \sum_j I_j$) forces
$c = -\kappa\, \bar g_R$. Rearranging gives $a_j = C_j(a)$. This interpretation uses
the same per-unit-of-capacity convention as the construction above.

\paragraph{The response function.} We choose $g$ increasing and convex in density,
so that the marginal penalty grows with density. The simplest one-parameter family
is $g(\tilde a) = \tilde a^{\rho}$ with $\rho > 1$; we use $\rho = 2$ throughout.

\paragraph{Consequences.} Conservation holds exactly at every state:
$\sum_j C_j = \sum_j I_j$. Subdivision invariance holds as well, and we verified it
numerically in two tests. Running a system against its everywhere-subdivided copy
for 20 iterations gives a maximum per-zone relative difference of at most
$1\times10^{-12}$. Running it to tolerance against a copy in which a single zone
is split gives $6\times10^{-16}$ between the converged states.

\paragraph{Domain.} For sufficiently large $\kappa$, some components of $C(a)$ can
become negative. The prescribed update is admissible only when $C_j(a) \ge 0$ for
every active zone. If this condition fails, the solver stops with a typed domain
error (Section~\ref{sec:failstop}).

\subsection{Scope}
\label{sec:notheres}

This note specifies the penalty and reports numerical evidence: the reproduction
of Section~\ref{sec:solver}, the examples of Section~\ref{sec:illustrations}, and
multi-start sweeps that count the distinct converged states found at each point of
two parameter grids (Section~\ref{sec:illustrations} and
Appendix~\ref{app:multistart}). The sweeps record which grid points had more than one
state under a stated protocol. Why several states occur, existence and uniqueness
results, analysis of the linearized dynamics, and closed-form characterizations are
the subject of companion work in preparation.

\section{Reference implementation}
\label{sec:solver}

\prefcent{} computes equilibria by the model's own adjustment dynamics: damped
fixed-point iteration $a \leftarrow (1-\omega)\,a + \omega\, C(a)$, followed by
renormalization of the total, each step. Because $C$ conserves the total,
renormalizing before or after the damping step gives the same trajectory in exact
arithmetic; the implementation renormalizes last.

The default initial state is $a = R$ and the default damping is $\omega = 1$
(direct substitution). These are the conventions of the Porto Alegre computation
(Section~\ref{sec:porto}).

The package depends only on NumPy and SciPy \citep{numpy2020,scipy2020}. Kernels
are supplied as dense matrices, built from cost matrices or edge-weighted graphs by
included helpers, or wrapped from any operator that provides matrix-vector
products.

The checks described below document numerical stopping and feasibility. They do
not prove existence, uniqueness or stability, or bound the distance to an exact
equilibrium.

\subsection{Stopping criteria}
\label{sec:budget}

The stopping test uses the normalized damped step
\begin{equation}
  s_t \;=\; \frac{\max_{j\,\text{active}} \bigl|a^{(t+1)}_j - a^{(t)}_j\bigr|}{\bar R},
  \qquad
  \bar R \;=\; \frac{1}{N_{\text{active}}} \sum_{j\,\text{active}} R_j ,
  \label{eq:step}
\end{equation}
evaluated on the completed (damped, renormalized) step. The run stops when
$s_t \le \text{tol}$ and returns $a^{(t+1)}$. Every reported step norm uses
equation~(\ref{eq:step}). In exact arithmetic, $s_t/\omega$ is the normalized
fixed-point residual at $a^{(t)}$. A residual at the returned state requires
another evaluation of $C$; the Figure~\ref{fig:rescue} checks use this extra
evaluation without advancing the iteration.

Runs that meet the stopping tolerance report \texttt{CONVERGED}. Runs configured
with a fixed iteration budget and no tolerance report \texttt{FIXED\_BUDGET},
together with the final step norm. The Porto Alegre reproduction uses exactly 200
iterations, matching the original protocol; further iteration would change the
reproduced outputs.

\subsection{Domain and feasibility checks}
\label{sec:failstop}

Domain violations raise typed errors containing the iteration number and the
offending state. The solver does not clip negative values, because clipping would
change the specified update.

Before the first step the solver probes the kernel's support. Zones that can
neither send nor receive (zero row and zero column of $K$) are \emph{isolated}: a
certificate lists them together with the capacity excluded from the equilibrium,
and the active set is their complement. A zone that could receive but not send
would have a zero denominator in equation~(\ref{eq:gravity}); the solver rejects
such a kernel with a typed domain error. A zone that sends but cannot receive is
active and carries no inflow. Every kernel in this note has symmetric support, so
the active set is the set of non-isolated zones, and the same set is used in every
sum of Section~\ref{sec:penalty}, in the stopping rule and in the feasibility
checks.

At the returned state, the solver reports the smallest component of $C(a)$ and the
fixed-state penalty limit
\begin{equation}
  \kappa^*(a) \;=\; \min_{j:\; g(\tilde a_j) > \bar g_R}\;
  \frac{I_j(a)}{R_j\,\bigl(g(\tilde a_j) - \bar g_R\bigr)},
  \label{eq:margin}
\end{equation}
the strength at which the first active zone's update would reach zero with $a$
held fixed; an empty minimum is reported as unbounded. The calculation holds $a$
fixed. It does not give a safe parameter range for a trajectory or for solutions
recomputed at other penalty strengths, and every iterate is checked separately. At
uniform density the penalty vanishes and the limit is unbounded. In
Section~\ref{sec:illustrations}, the penalty strengths were chosen as fractions of
the unpenalized state's limit.

\subsection{Reproducibility records}
\label{sec:provenance}

Every run can emit a manifest: a record of the run's inputs, settings and outputs,
in two parts. The fingerprint records what determines the result: the kernel's
identity and, when built by the package, the hash of the graph it came from; the
hash of the capacity vector; the initial state's convention and hash; the model and
solver parameters, including the stopping criterion by name, $\omega$, the
tolerance and the budget; and the numerical environment. The observations record
what happened: status, iterations, final step norm, certificates, statistics, and
a hash of the result.

Bit-for-bit repeatability on one platform requires a content-hashed kernel (or an
operator identity its supplier marks as trusted), a clean release or commit as the
source identity, a known BLAS threading configuration, and deterministic kernel
operators. The software records whether these requirements are met. When one
cannot be established, the manifest labels the run as not reproducible and lists
the reasons; the label means that the guarantee is unavailable, not that a rerun
produced different values.

The network, the capacity vector and the initial state are recorded separately.
This makes it possible to identify which input changed between experiments.

\subsection{Reproducing the Porto Alegre results}
\label{sec:porto}

We tested the implementation by reproducing the outputs released with the 2025
preprint of a companion study \citep{portoalegre}, which is under review: two
preferential centrality outputs, the study's PC1 ($\gamma = 1$) and PC2
($\gamma = 0$). The study's published data\footnote{%
\url{https://github.com/alexanderhellervik/porto-alegre-paper}: the segment file
\texttt{edges\_combined.shp} ($65{,}357$ segments), with the two outputs in columns
\texttt{cd1g1b2k0} ($\gamma = 1$) and \texttt{cd4g0b2k0} ($\gamma = 0$).}
contain its network, $65{,}357$ segments of a metropolitan network that extends
beyond the municipality, and the two outputs. The code that produced them ran 200
iterations of direct substitution from $a = R$. From that network and capacities, \prefcent{} rebuilds the routing graph, the cost matrix and
both outputs at $\kappa = 0$; the study's outputs are read only for comparison. All
$65{,}357$ segments agree with the study's outputs to within $0.1\%$. The median
relative difference is $2.2\times10^{-14}$ for $\gamma = 1$ and
$8.9\times10^{-15}$ for $\gamma = 0$, with a maximum of $1.6\times10^{-12}$. The
rebuilt cost matrix agrees with the archived one bit for bit. The original
computation was made in 2024 with a different linear algebra stack. Details and the
export contract are in the package's \texttt{REPRODUCE.md}. The package's
regression tests also compare the update map with archived reference trajectories,
to $3\times10^{-15}$.

\section{Illustrations}
\label{sec:illustrations}

All synthetic runs use equation~(\ref{eq:gravity}) with $\beta = 2$, $d_0 = 5000$
and no self-interaction. Costs are measured in meter-equivalents at 60~km/h, so
$d_0$ corresponds to a five-minute access cost. The response function is
$g(\tilde a) = \tilde a^2$.

We use two capacity landscapes: a 512-zone ring and a $64\times64$ torus, a
square grid with opposite edges connected. Zones are 1~km apart. The ring has a
spatially correlated lognormal capacity landscape, with a log-standard deviation
of $0.5$ and a correlation length of 10~km. The torus has four designed capacity
hills, local maxima of the capacity landscape, roughly 30~km apart. Both
landscapes are scaled to mean capacity $1$, so the normalized step in
equation~(\ref{eq:step}) equals the raw maximum step.

\begin{figure}[htbp]
  \centering
  \includegraphics[width=\linewidth]{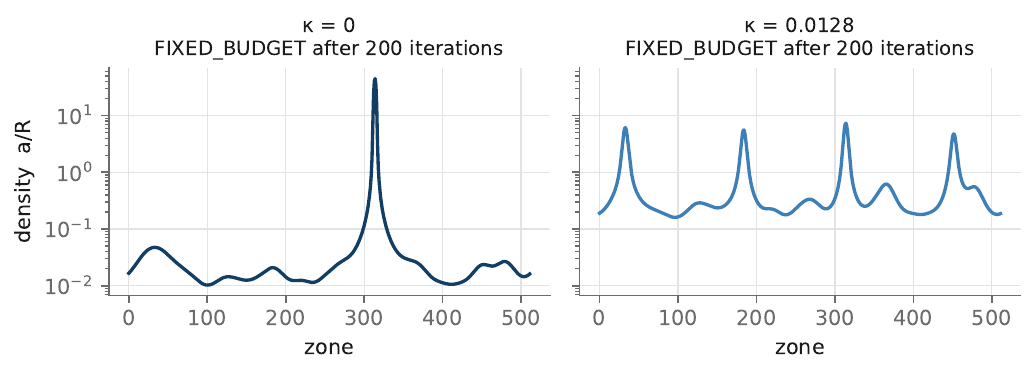}
  \caption{\textbf{Concentration after 200 iterations.} Both runs use the ring
  with $\gamma = 2$, $\omega = 1$ and initial state $a = R$. Without the penalty,
  the maximum density is $44.9$ and $97\%$ of activity lies within 10~km of the
  largest peak. With $\kappa = 0.0128$, the maximum density is $7.3$ and four
  peaks pass the detection rule. The normalized step norms are
  $9.6\times10^{-2}$ and $1.6\times10^{-3}$, respectively. Both panels show
  \texttt{FIXED\_BUDGET} outputs. Continuing the runs to the $10^{-12}$ tolerance
  requires 7,196 iterations without the penalty and 1,607 with it.}
  \label{fig:rescue}
\end{figure}

\begin{figure}[htbp]
  \centering
  \includegraphics[width=\linewidth]{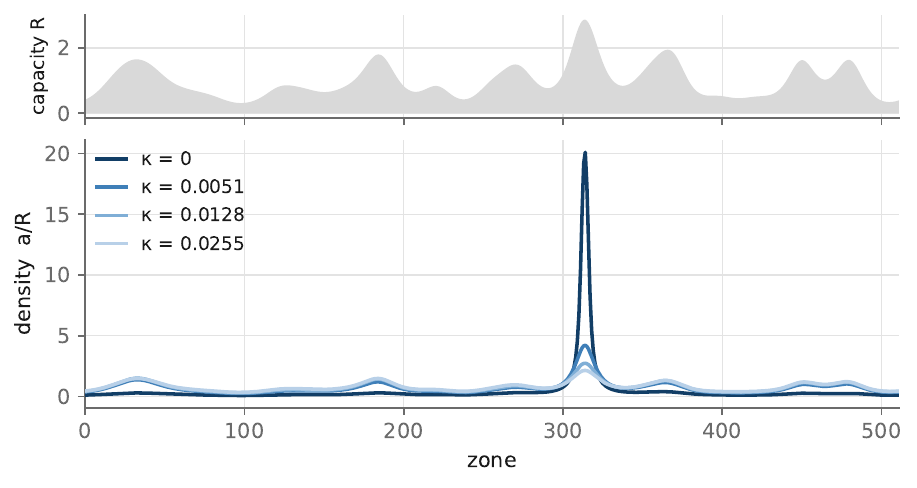}
  \caption{\textbf{Penalty strength and peak density.} On the ring at
  $\gamma = 0.5$, the four penalty strengths are $\kappa = 0$, $0.0051$, $0.0128$
  and $0.0255$. The corresponding peak densities are 20, $4.2$, $2.7$ and $2.2$,
  and the detected peak counts are one, two, four and four. The positive penalty
  strengths are approximately $10\%$, $25\%$ and $50\%$ of the unpenalized
  state's limit, $\kappa^* = 0.051$.}
  \label{fig:series}
\end{figure}

\begin{figure}[htbp]
  \centering
  \includegraphics[width=\linewidth]{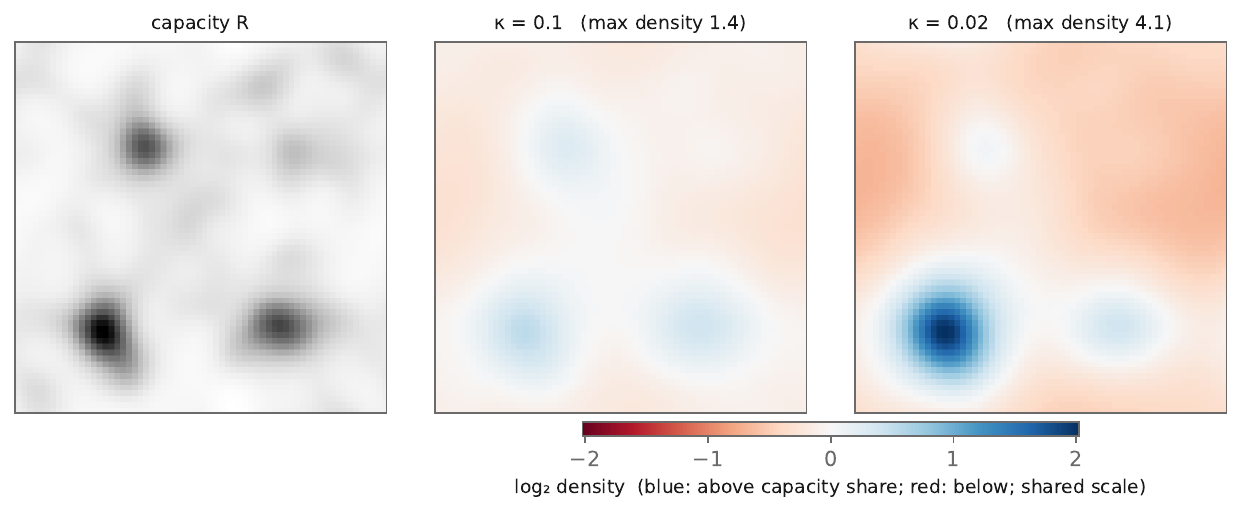}
  \caption{\textbf{Density contrasts across the four-hill landscape.} On the torus
  at $\gamma = 1.5$, the larger penalty reduces differences in density among the
  capacity hills. Maximum density is $1.44$ for $\kappa = 0.1$ and $4.06$ for
  $\kappa = 0.02$; the corresponding minimum densities are $0.80$ and $0.62$.
  Both states have two detected centers under the rule of
  Section~\ref{sec:illustrations}. The maps show $\log_2(a/R)$ on a shared scale:
  blue for activity above the capacity share and red for activity below it.}
  \label{fig:torus}
\end{figure}

\begin{figure}[htbp]
  \centering
  \includegraphics[width=\linewidth]{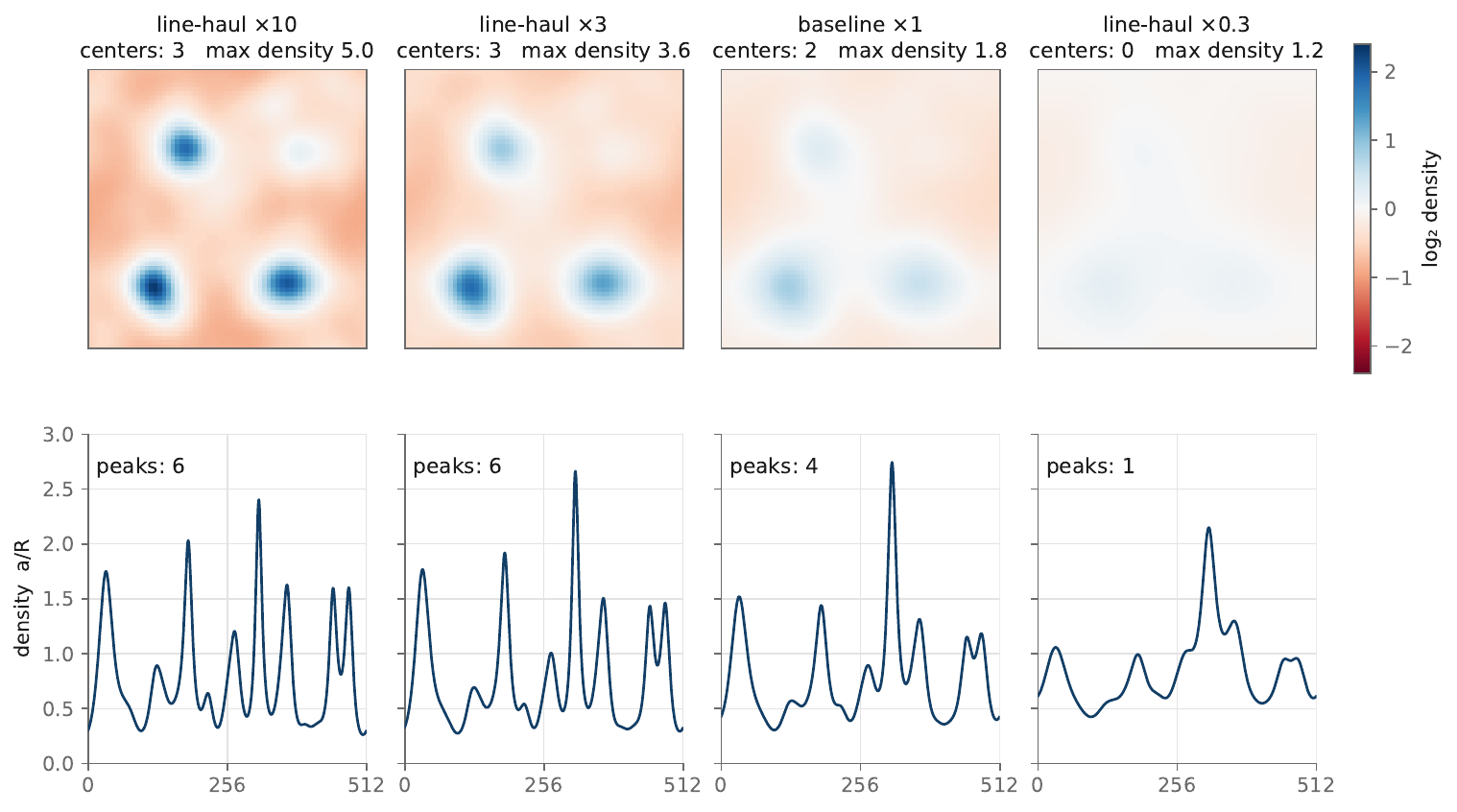}
  \caption{\textbf{Transport cost and detected centers.} Line-haul costs are
  multiplied by $s = 10$, $3$, $1$ and $0.3$, while $d_0$ remains fixed. The torus
  uses $\gamma = 1.5$ and $\kappa = 0.05$; the ring uses $\gamma = 0.5$ and
  $\kappa = 0.0128$. Across these settings, lower costs are associated with fewer
  detected centers. The torus counts are three, three, two and zero; the ring
  counts are six, six, four and one. Zero centers does not mean uniform density:
  the torus maximum at $s = 0.3$ is $1.16$. The largest ring peak occurs at
  $s = 1$. These panels compare separately computed equilibria.}
  \label{fig:technology}
\end{figure}

\begin{figure}[htbp]
  \centering
  \includegraphics[width=\linewidth]{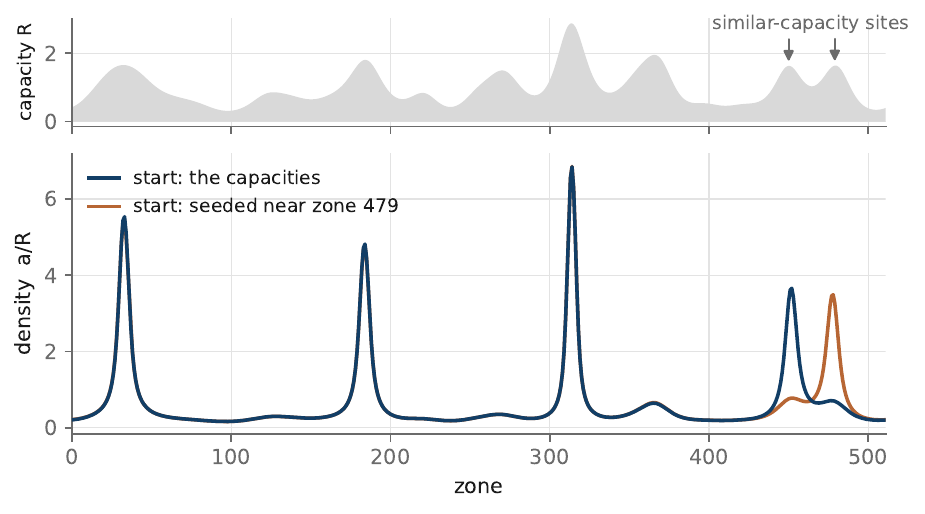}
  \caption{\textbf{One landscape, two equilibria.} The ring uses $\gamma = 1.75$
  and $\kappa = 0.0128$. One run starts from $a = R$; the other starts from a
  distribution seeded near the capacity hill at zone 479. All other inputs are
  fixed. The runs reach different numerical equilibria, with three centers in the
  same locations and the fourth near zone 452 or zone 478. Both meet the
  $10^{-12}$ stopping tolerance and pass the feasibility checks. Each is recovered
  after the $10^{-4}$ multiplicative perturbation test, with renormalization. The
  manifests record the difference in initial states; the test protocol is in the
  archive.}
  \label{fig:siting}
\end{figure}

\begin{figure}[htbp]
  \centering
  \includegraphics[width=\linewidth]{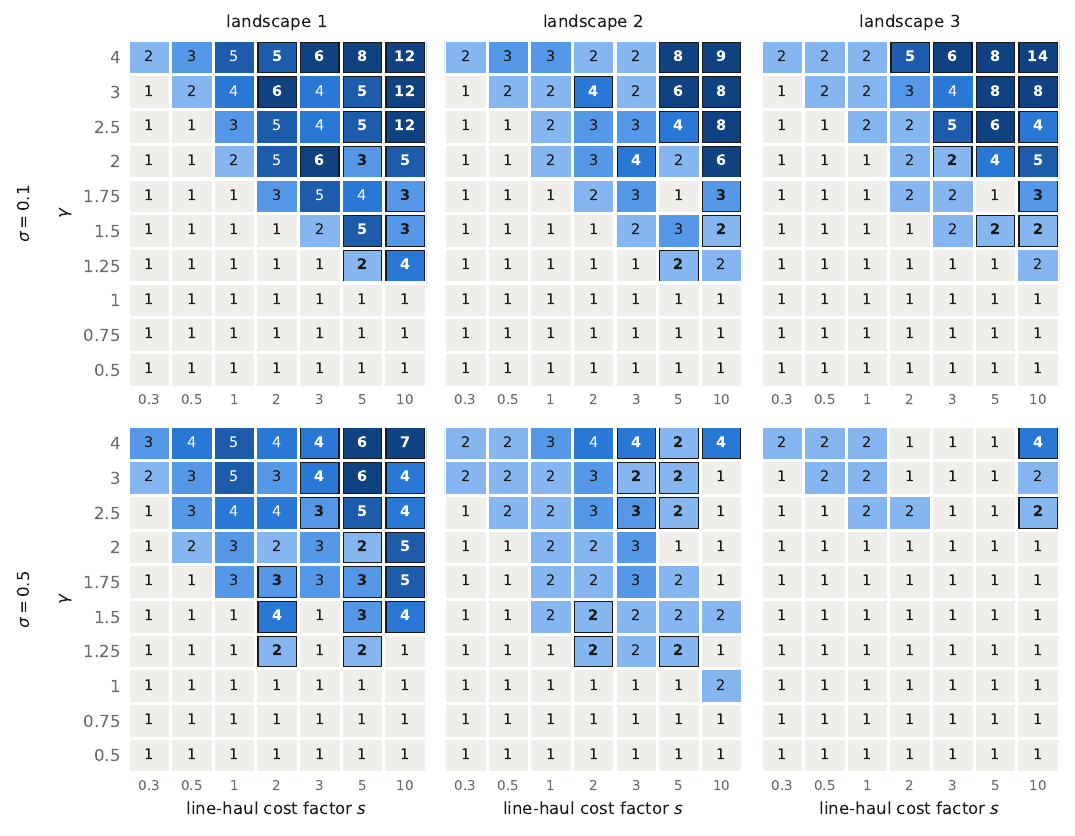}
  \caption{\textbf{Distinct converged states across a parameter grid.} Torus,
  three correlated lognormal landscapes (columns) at capacity log-standard
  deviation $\sigma = 0.1$ (top) and $0.5$ (bottom), $\kappa = 0.01$,
  $\omega = 0.5$. Each cell gives the number of distinct converged states reached
  from the same 16 initial states at that $\gamma$ and line-haul cost factor $s$;
  it is a lower bound. Outlined cells in bold: the states differ in their number of
  detected centers. All 6{,}720 runs met the $10^{-12}$ tolerance and passed the
  feasibility checks, and every distinct state returned to itself after the
  perturbation test of Figure~\ref{fig:siting}.}
  \label{fig:multistart}
\end{figure}

Except for the 200-iteration comparison in Figure~\ref{fig:rescue}, every run
behind Figures~\ref{fig:rescue}--\ref{fig:multistart} met the $10^{-12}$ stopping
tolerance and passed the feasibility checks; Appendix~\ref{app:multistart} reports
the sweep runs that did not. Table~\ref{tab:runs} summarizes the principal runs. The archived scripts
give seeds and remaining implementation details.

We count local density maxima above $\tilde a = 1.3$. On the ring, a peak must be
strictly higher than its immediate neighbors. On the torus, maxima are detected in
$9\times9$ neighborhoods that wrap around the periodic edges. A count of zero means that no maximum passes this rule.

\begin{table}[htbp]
  \centering\small
  \caption{Principal runs behind the figures. The final step is the normalized
  damped step $s_t$ of equation~(\ref{eq:step}). $\kappa^*$ is the fixed-state
  penalty limit of equation~(\ref{eq:margin}) at the returned state; the
  feasibility checks pass for every run. Full records are in the archived
  manifests. The Figure~\ref{fig:multistart} row summarizes 6{,}720 runs; the
  sweeps of Appendix~\ref{app:multistart} are listed in Table~\ref{tab:sweeps}.}
  \label{tab:runs}
  \setlength{\tabcolsep}{5pt}
  \begin{tabular}{llllllll}
    \hline
    Figure & $\gamma$ & $\kappa$ & $\omega$ & status & iterations & final step & $\kappa^*$ \\
    \hline
    1 & 2 & $0,\ 0.0128$ & 1 & \texttt{FIXED\_BUDGET} & 200 & $0.096,\ 0.0016$ & $0.023,\ 0.16$ \\
    2 & 0.5 & $0\ldots0.0255$ & 0.5 & \texttt{CONVERGED} & 560--1722 & $\le 10^{-12}$ & $0.051$--$0.65$ \\
    3 & 1.5 & $0.02,\ 0.1$ & 0.5 & \texttt{CONVERGED} & 194--2431 & $\le 10^{-12}$ & $0.29,\ 1.5$ \\
    4, torus & 1.5 & 0.05 & 0.5 & \texttt{CONVERGED} & 184--551 & $\le 10^{-12}$ & $0.26$--$3.4$ \\
    4, ring & 0.5 & 0.0128 & 0.5 & \texttt{CONVERGED} & 625--1193 & $\le 10^{-12}$ & $0.45$--$0.63$ \\
    5 & 1.75 & 0.0128 & 0.5 & \texttt{CONVERGED} & 2542--2622 & $\le 10^{-12}$ & $0.17,\ 0.17$ \\
    6 & 0.5--4 & 0.01 & 0.5 & \texttt{CONVERGED} & 88--344{,}040 & $\le 10^{-12}$ & $0.032$--$18$ \\
    \hline
  \end{tabular}
\end{table}

\paragraph{Concentration at a practical budget.} Figure~\ref{fig:rescue} asks what
the penalty changes at a practical iteration budget. At $\gamma = 2$ on the ring,
both runs use the model's own map ($\omega = 1$) from $a = R$ for 200 iterations.
Without the penalty, concentration is still increasing at the budget. Continued at
the same settings, the run needs 7,196 iterations to meet the tolerance, at a state
with $98\%$ of activity within 10~km of the peak. With $\kappa = 0.0128$ the step
norm is $4.7\times10^{-9}$ at 1,000 iterations and the tolerance is met at 1,607.
For each returned state we also evaluated the residual at that state by one extra
evaluation of $C$; it differs from the reported step by at most $7\%$.

\paragraph{Penalty strength.} Figure~\ref{fig:series} varies the penalty strength
on the ring at $\gamma = 0.5$. The unpenalized equilibrium concentrates on the
landscape's strongest region. With the three positive strengths, the peak density
falls and the number of detected peaks rises from one to four. The peaks remain
associated with the capacity landscape.

\paragraph{Density contrasts among centers.} Figure~\ref{fig:torus} compares two
penalty strengths on the four-hill torus at $\gamma = 1.5$; the hill capacities
are $3.0$, $2.4$, $2.2$ and $1.4$ times the mean. In the order of the hills'
capacities, the hill-associated density maxima are $1.44$, $1.31$, $1.20$ and
$0.96$ at $\kappa = 0.1$, and $4.06$, $1.31$, $1.04$ and $0.73$ at
$\kappa = 0.02$. The larger penalty reduces the differences in density among the
hills. It does not change the count of detected centers, which is two in both
states.

\paragraph{Transport cost.} Figure~\ref{fig:technology} varies transport cost. We
represent a change in transport conditions by scaling all line-haul costs by a
factor $s$ while holding the fixed access cost $d_0$ constant. Scaling both
together would multiply the kernel by a common factor, which cancels in the
normalized interaction shares of equation~(\ref{eq:gravity}). Across the four
factors tested, on both landscapes, lower line-haul costs are associated with fewer
detected centers. The ring's maximum density is $2.40$, $2.66$, $2.74$ and $2.15$
for $s = 10$, $3$, $1$ and $0.3$: the largest peak is highest at the baseline and
lower at both slower and faster transport. The panels compare separately computed
equilibria and do not show a merger process.

\paragraph{Initial state.} Figure~\ref{fig:siting} shows that the initial state can
select among equilibria. For each earlier parameter setting, the initial states
tested to convergence reached the same numerical state. On the ring at
$\gamma = 1.75$ and $\kappa = 0.0128$, the landscape contains two capacity hills
of similar height at zones 450 and 479, 29~km apart, with capacities $1.63$ and
$1.64$. A run started from $a = R$ and a run started from a state seeded near
zone 479 reach different equilibria, with three shared centers and the fourth at
zone 452 (density $3.7$) or zone 478 (density $3.5$). Each returns to itself after
a multiplicative perturbation of relative size $10^{-4}$ with renormalization, to
within $7\times10^{-14}$ and $4\times10^{-11}$ relative. The two manifests are
identical in all input fields except the initial state.

\paragraph{Initial states across a parameter grid.} Figure~\ref{fig:multistart}
asks how common the situation of Figure~\ref{fig:siting} is. On the torus we use
three further capacity landscapes: spatially correlated lognormal fields with a
correlation length of 4~km and a log-standard deviation $\sigma$ of $0.1$ or $0.5$,
scaled to mean capacity $1$. At each point of a grid in $\gamma$ (from $0.5$ to $4$)
and in the line-haul cost factor $s$ of Figure~\ref{fig:technology} (from $0.3$ to
$10$), with $\kappa = 0.01$, the model is run from the same 16 initial states, and
we count the distinct converged states. Two converged states count as the same when
they differ by at most $10^{-6}$ in the norm of equation~(\ref{eq:step}). The
observed differences were at most $5\times10^{-9}$ within a state and at least $10$
between states, so the count does not depend on this choice. Each distinct state
was perturbed as in Figure~\ref{fig:siting} and returned to itself.

More than one state was found at 172 of the 420 grid points, on all three
landscapes; at 75 of them, the states differ in their number of detected centers.
All but one of these points have $\gamma \ge 1.25$, and they are more frequent at
larger $\gamma$ and larger $s$. The counts are lower bounds: at $\gamma = 4$ and
$s = 10$, up to 14 of the 16 initial states reached different states.
Appendix~\ref{app:multistart} gives the protocol, a second grid in $\gamma$ and the
capacity amplitude $\sigma$, an example pair of states, and every sweep that was
run. The sweeps cover one lattice, one response function and fixed penalty
strengths. They record where distinct states were found under this protocol; they
do not explain why.

\section{Using historical inputs}
\label{sec:enables}

Historical conditions can enter the model through three inputs: the transport
network, the capacity landscape and the initial distribution of activity. Each can be
varied separately: an earlier network can be combined with current capacities, or an
earlier capacity landscape with the current network. A sequence of networks can be
evaluated by using each computed state as the initial state for the next network; the
run records identify the inputs of each step.

The initial distribution matters because the model need not have a single
equilibrium. Figure~\ref{fig:siting} shows two numerical equilibria on one network and
capacity landscape, reached from different initial states, and
Figure~\ref{fig:multistart} shows that this happens over a range of parameter values.
Where it happens, the state the model reaches depends on where activity was before,
not only on the current network and capacities. An empirical test would start one run
from a historical activity distribution and one from the default $a = R$, and compare
both with the observed pattern; Figure~\ref{fig:technology} illustrates the
corresponding comparison for transport costs.

Whether these mechanisms explain observed settlement patterns remains an empirical
question. This note provides the model specification and the computational workflow
for such tests.

\section{Software availability}
\label{sec:software}

\prefcent{} version 0.1.1 is released under the MIT license. Its runtime
dependencies are NumPy and SciPy. Source code, tests and reproduction instructions
(\texttt{REPRODUCE.md}) are at \url{https://github.com/prefcent/prefcent};
releases are on PyPI and archived at Zenodo (concept DOI
\texttt{10.5281/zenodo.22643613} for all versions). The release used for this note
is identified by its version DOI \texttt{10.5281/zenodo.23103539}.

The figures are reproduced from a separate archive, the repository
\texttt{density-penalty-paper}\footnote{\url{https://github.com/prefcent/density-penalty-paper}}
(version DOI \texttt{10.5281/zenodo.23158476}), which holds the run scripts,
the archived state and manifest of every run behind Figures~\ref{fig:rescue}--\ref{fig:siting},
the per-figure summaries, and the rendering script. For the sweeps of
Figure~\ref{fig:multistart} and Appendix~\ref{app:multistart} it holds the manifest
of every run, the state of every distinct equilibrium found, and the counting
script. Its check script re-solves runs and compares the result hashes with the
archived manifests.

Citation: this note and \citet{hellervik2019}. \texttt{CITATION.cff} carries both.

\section*{Use of AI tools}
\label{sec:ai}

AI tools were used extensively in this work. Claude models (Anthropic; Claude
Opus~4.8, Opus~5, Opus~5.5 and Fable~5), working as coding agents through Claude Code,
wrote most of the \prefcent{} source code and tests to the author's specification,
ran the computations, built the reproduction archive and rendered the figures, drafted
and revised the text of this note, and checked each cited source against a local copy
of it. Cursor's coding agent implemented parts of the first solver milestones, and
OpenAI Codex implemented fixes found in a release review. OpenAI GPT-6 models reviewed
drafts of the note and of the software; the author decided which findings to adopt.
The author formulated the model and the density penalty, made the design and
scientific decisions, and reviewed all of the \prefcent{} source code and all of the
scripts in the reproduction archive. The author has reviewed, and takes
responsibility for, all content of this note.

\bibliographystyle{plainnat}
\bibliography{refs-display}

\appendix
\section{Multi-start protocol and further grids}
\label{app:multistart}

\paragraph{Landscapes.} Each of the three landscapes is a fixed pattern: white noise
on the $64\times64$ torus, smoothed with a periodic Gaussian filter of width 4~km and
standardized. Capacity is $R_j \propto \exp(\sigma z_j)$ for pattern values $z_j$,
scaled to mean $1$, so changing $\sigma$ changes the amplitude of the landscape but
not its pattern. Costs, $\beta$, $d_0$, $\rho$ and the absence of self-interaction
are those of Section~\ref{sec:illustrations}.

\paragraph{Initial states.} The same 16 initial states are used at every grid point
of a landscape: $a = R$; uniform activity; seven states $R_j \exp(w_j)$ with $w$ a
correlated Gaussian field (width 3~km, unit standard deviation); and seven states
$R_j\,(1 + 4 \sum_k b_{kj})$ with two to five Gaussian bumps $b_k$ (width 3~km) at
random sites. All are rescaled to total activity $\sum_j R_j$. Seeds are fixed and
recorded in the archive; each manifest records the hash of its initial state.

\paragraph{Runs and counting.} Every run uses $\omega = 0.5$, the $10^{-12}$
tolerance and a budget of $400{,}000$ iterations. The torus kernel is applied by fast
Fourier transform through the package's operator interface
(Section~\ref{sec:solver}); against the dense kernel it agrees to $10^{-15}$ per
matrix--vector product, and a full test run reaches the same state to within
$6\times10^{-14}$ in the same number of iterations. At each grid point, converged
states are grouped greedily: a state joins an earlier group if it lies within
$10^{-6}$ of that group's first state in the norm of equation~(\ref{eq:step}).
Across both grids, states in one group differ by at most $4\times10^{-8}$ and
states in different groups by at least $10$. Starts that exhaust the budget or stop
with a domain error are reported and not counted. Each distinct state is perturbed
multiplicatively with relative size $10^{-4}$, renormalized and run to tolerance
again; all 1{,}110 distinct states returned to within $4\times10^{-8}$ of
themselves.

\begin{figure}[htbp]
  \centering
  \includegraphics[width=0.92\linewidth]{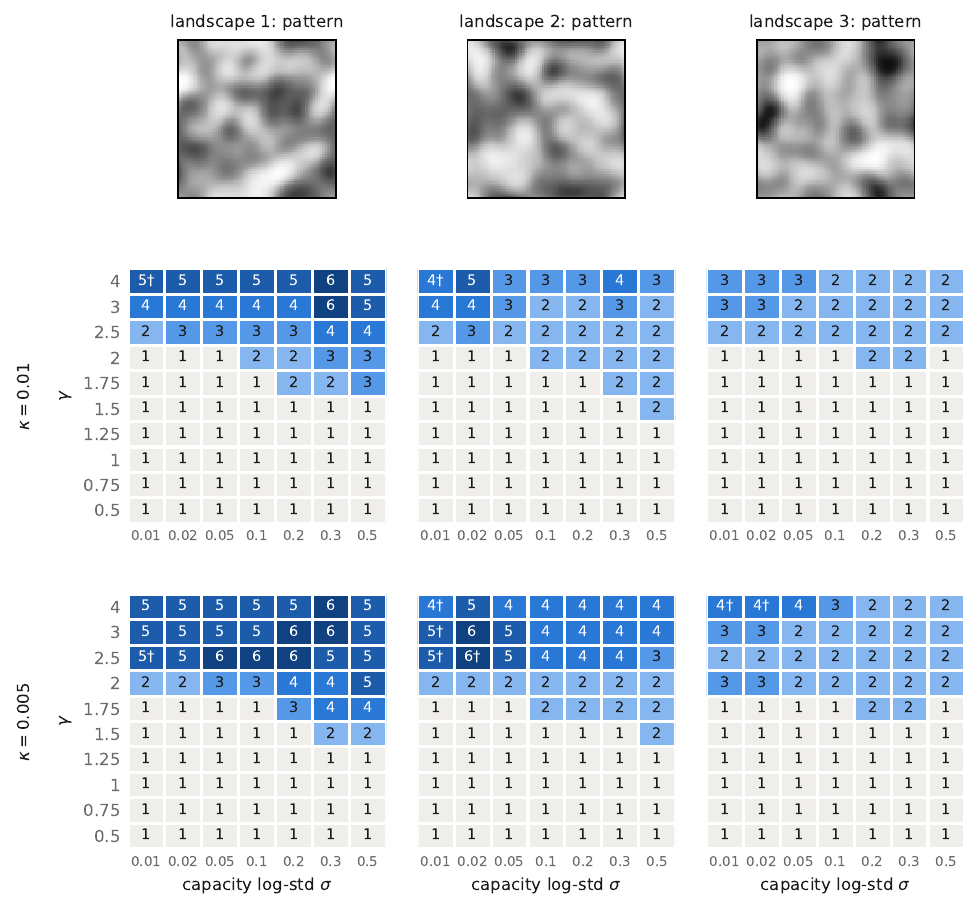}
  \caption{\textbf{Distinct converged states against capacity amplitude.} Top:
  the three capacity patterns. Below: number of distinct converged states reached
  from 16 initial states at each $\gamma$ and capacity log-standard deviation
  $\sigma$, with $s = 1$ and $\kappa = 0.01$ (middle) or $0.005$ (bottom); a lower
  bound. $\dagger$: at least one start exhausted the iteration budget and is not
  counted.}
  \label{fig:amplitude}
\end{figure}

\begin{figure}[htbp]
  \centering
  \includegraphics[width=0.8\linewidth]{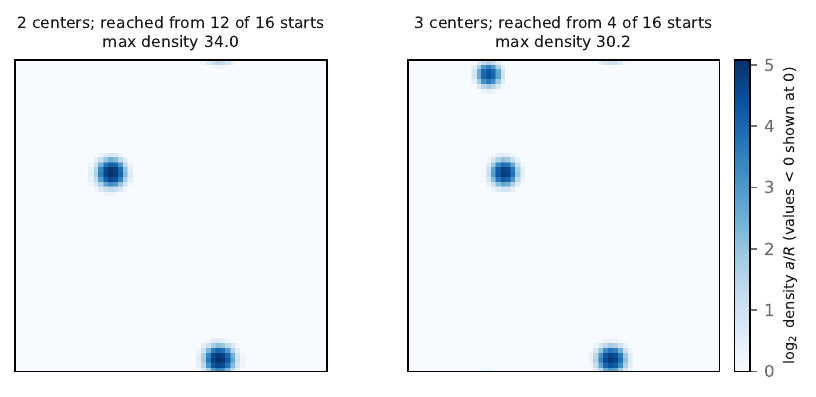}
  \caption{\textbf{Same inputs, two states.} Landscape 2 of
  Figure~\ref{fig:multistart} at $\sigma = 0.5$, $\gamma = 3$, $\kappa = 0.01$,
  $s = 3$. Twelve of the 16 initial states reach the state with two centers (left)
  and four reach the state with three (right). Density is shown as $\log_2(a/R)$;
  centers crossing the periodic edges appear at both sides of a panel.}
  \label{fig:pair}
\end{figure}

\paragraph{Capacity amplitude.} Figure~\ref{fig:amplitude} holds $s = 1$ and varies
$\gamma$ and $\sigma$ at $\kappa = 0.01$ and $0.005$. More than one state was found
at 79 and 96 of the 210 grid points, respectively, on all three landscapes, and
only at $\gamma \ge 1.5$. There is no single trend in $\sigma$: at
$\gamma = 1.75$--$2$, distinct states are found mainly on the stronger landscapes,
while at $\gamma \ge 2.5$ some landscapes give more states when $\sigma$ is small. At
$s = 1$, 560 of the 567 distinct states have a single detected center and differ in
its location. At seven points ($\gamma = 2$, $\kappa = 0.005$, $\sigma \le 0.05$),
one of the states found has no center that passes the detection rule. Of the
$6{,}720$ runs, 25 exhausted the budget, all at $\sigma \le 0.02$ and $\gamma \ge 2.5$. Figure~\ref{fig:pair} shows an example with
different numbers of centers from the grid of Figure~\ref{fig:multistart}.

\paragraph{All sweeps.} Table~\ref{tab:sweeps} lists every sweep run for this note,
including the preliminary grids that preceded the design of
Figures~\ref{fig:multistart} and~\ref{fig:amplitude}. The preliminary $\gamma$--$s$
grids use the three landscapes at $\sigma = 0.5$ and the four-hill landscape of
Figure~\ref{fig:torus}, with eight initial states (the first eight of the 16). At
$\kappa = 0.05$, 62 starts stopped with a domain error before converging, all from
the correlated random initial states and all at $\gamma \le 1.75$ and $s \le 3$;
no point had more than one state.

\begin{table}[htbp]
  \centering\small
  \caption{Every multi-start sweep run for this note. ``Points'' are grid points;
  ``$\ge 2$'' counts points with more than one distinct converged state. Budget and
  domain columns count starts that were not counted.}
  \label{tab:sweeps}
  \setlength{\tabcolsep}{4pt}
  \begin{tabular}{llllrrrrr}
    \hline
    Sweep & $\gamma$ & $s$ & $\sigma$ & $\kappa$ & starts & points & $\ge 2$ & budget / domain \\
    \hline
    preliminary & 0.5--2 & 0.3--10 & 0.5, hills & 0.05 & 8 & 196 & 0 & 0 / 62 \\
    preliminary & 0.5--2 & 0.3--10 & 0.5, hills & 0.02 & 8 & 196 & 4 & 0 / 0 \\
    preliminary & 0.5--2 & 0.3--10 & 0.5, hills & 0.01 & 8 & 196 & 27 & 0 / 0 \\
    amplitude & 0.5--2 & 1 & 0.01--0.2 & 0.02 & 12 & 105 & 0 & 0 / 0 \\
    probe & 2.5--4 & 1, 3 & 0.01, 0.05 & 0.01, 0.005 & 4 & 72 & 68 & 16 / 0 \\
    Figure~\ref{fig:amplitude} & 0.5--4 & 1 & 0.01--0.5 & 0.01 & 16 & 210 & 79 & 5 / 0 \\
    Figure~\ref{fig:amplitude} & 0.5--4 & 1 & 0.01--0.5 & 0.005 & 16 & 210 & 96 & 20 / 0 \\
    Figure~\ref{fig:multistart} & 0.5--4 & 0.3--10 & 0.1, 0.5 & 0.01 & 16 & 420 & 172 & 0 / 0 \\
    \hline
  \end{tabular}
\end{table}

\end{document}